\documentclass{PoS}
\usepackage{amssymb}
\usepackage{amsmath}
\usepackage{bm}
\usepackage{wrapfig}
\usepackage{graphicx}
\usepackage{braket}
\title{Spontaneous CP violation in the NJL model at $\theta = \pi$ }

\ShortTitle{Spontaneous CP-violation in the NJL model at $\theta = \pi$ }

\author{\speaker{Jorn K. Boomsma}\\
        Vrije Universiteit Amsterdam\\
        E-mail: \email{jboomsma@few.vu.nl}}

\author{Dani\"el Boer\\
        Vrije Universiteit Amsterdam\\
        E-mail: \email{dboer@few.vu.nl}}

\abstract{As is well-known, spontaneous CP-violation in the strong interaction
  is possible at $\theta = \pi$, which is commonly referred to as
  Dashen's phenomenon. This phenomenon has been studied extensively
  using chiral Lagrangians. 
  Here the two-flavor NJL model at $\theta = \pi$ is discussed. It
  turns out that the occurrence of spontaneous CP-violation depends on
  the strength of the 't Hooft determinant interaction, which
  describes the effect of instanton interactions.  The dependence of
  the phase structure, and in particular of the CP-violating phase, on
  the quark masses, temperature, baryon and isospin chemical potential
  is examined in detail. The latter dependence shows a modification of
  the charged pion condensed phase first discussed by Son and
  Stephanov.}

\FullConference{8th Conference Quark Confinement and the Hadron Spectrum\\
		 September 1-6, 2008\\
		 Mainz. Germany}

\begin{document}

\section{Introduction}
The study of spontaneous CP violation (SCPV) in the strong interaction
at $\theta = \pi$ has a long history.  Already in 1971
Dashen~\cite{Dashen:1970et} pointed out its possibility and it has
been studied extensively using chiral Lagrangians, see for example
Refs.~\cite{ChPT,Creutz:2003xu,Tytgat:1999yx}.  The angle $\theta$
stands for the vacuum angle that enters the QCD Lagrangian through the
term $\mathcal{L}_\theta = \frac{\theta g^2}{32 \pi^2} F \tilde F$ to
which instantons contribute. This $\theta$-term leads to explicit CP
violation, except when $\theta = 0~(\rm{mod}\ \pi)$. When $\theta =
\pi$ it may happen that while the Lagrangian is invariant under CP,
the ground state is not.

We will show that the actual occurrence of SCPV depends on the
strength of the instanton interaction. This will be discussed for the
two-flavor Nambu-Jona-Lasinio (NJL) model \cite{NJL}. In models like
the NJL-model and in low energy effective theories of the strong
interaction in general, the effects of instantons are mimicked by an
effective interaction, the 't Hooft determinant
interaction~\cite{tHooft}. We will investigate how SCPV depends on the
strength of this interaction.

Other properties of the SCPV phase are studied by calculating its
dependence on temperature and nonzero baryon and isospin chemical
potential. This could be of relevance to heavy-ion collisions, despite
the fact that in Nature $\theta < 10^{-10}$. Namely, it has been
suggested that metastable CP-violating bubbles might be created in
such collisions.  These states would be characterized by an effective,
possibly large $\theta$, cf.\ Refs.~\cite{heavy_ion}. Experimental
signatures for these bubbles have been discussed in
Refs.~\cite{exp_sign}.

Finally, we discuss mixing of mesons with their
parity partners, which arises whenever CP invariance is broken. This affects 
the charged pion condensed phase that arises at sufficiently large 
isospin chemical potential. The results presented here are based on 
Ref.~\cite{Boer:2008ct} to which we refer for details. 

\section{The NJL model}
The following form of the 2-flavor NJL-model is used
\begin{eqnarray}
 \mathcal{L} = \bar \psi \left(i \gamma^\mu \partial_\mu + \gamma_0
 \mu \right) \psi - \bar \psi M_0 \psi + G_1 \left[ (\bar \psi
   \lambda_a \psi)^2 + (\bar \psi \lambda_a i \gamma_5 \psi)^2 \right]
 + \nonumber \\ G_2 \left[ e^{i \theta} \det \left( \bar \psi_R \psi_L
   \right) + \mathrm{h.c.} \right].
\end{eqnarray}
The interaction with coupling constant $G_1$ is chirally symmetric. The
interaction proportional to $G_2$ is the 't Hooft determinant
interaction and represents the effects of instantons. As the model is
non-renormalizable, a cut-off is needed. Here a three-dimensional
cut-off is used, which is set by the value of  
the chiral condensate $\braket{\bar \psi \psi}$, via a gap equation. 
The coupling constants depend on this
cut-off as $G_i \sim \mathcal{O}(1) / \Lambda^2$ for dimensional reasons.
We restrict to two flavors, using $\lambda_a$ with $a=0,...,3$ as
the generators of U(2).

\begin{figure}[t]
   \begin{minipage}[t]{0.45\linewidth}
     \centering
     \includegraphics[width=\textwidth]{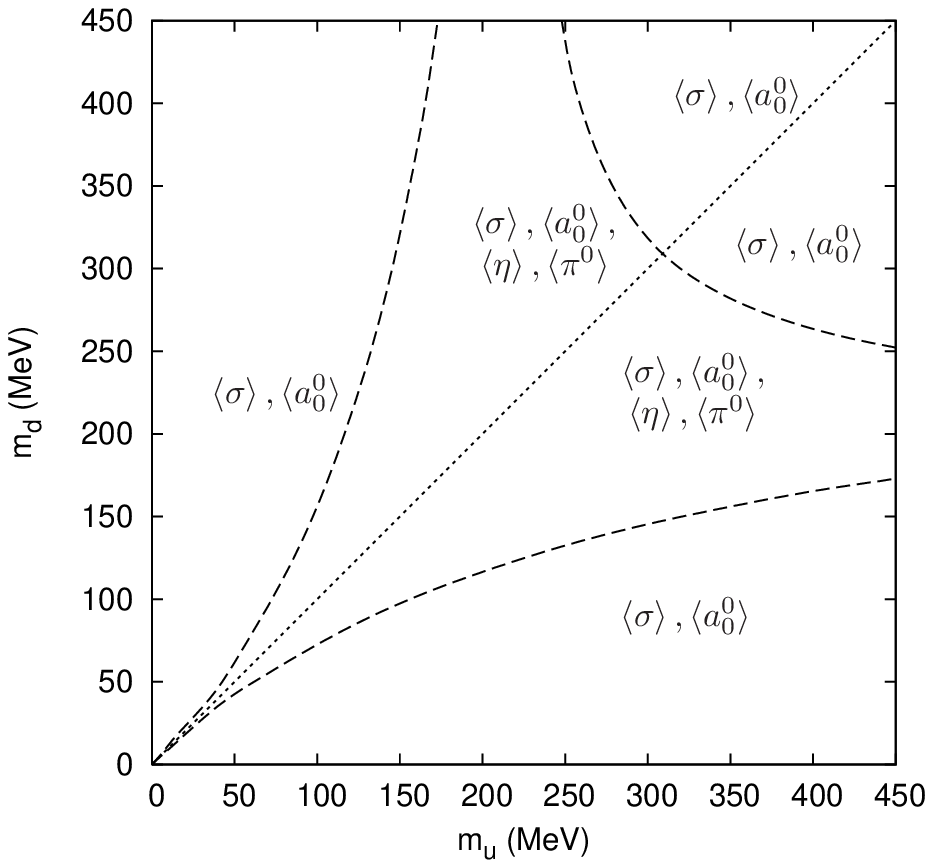}
     \caption{$(m_d,m_u)$ phase diagram with $c =0.4$.}
     \label{phasediagram_mu0_md0}
   \end{minipage}
   \hspace{0.1\linewidth}
   \begin{minipage}[t]{0.45\linewidth}
     \centering
     \includegraphics[width=\textwidth]{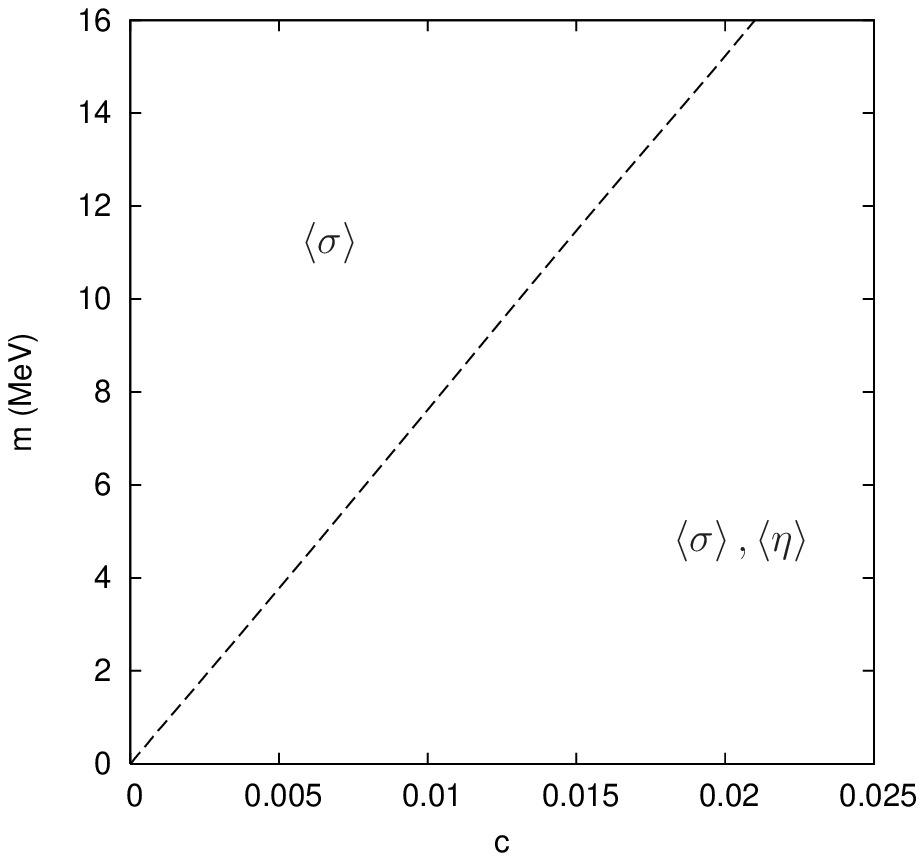}
     \caption{$(m, c)$ phase diagram.}
     \label{phasediagram_m_c}
   \end{minipage}
\end{figure}

We wish to investigate the dependence of SCPV at $\theta = \pi$ on the
strength of the instanton interaction $G_2$, while keeping the physics
at $\theta = 0$ unchanged. This means that the sum $G_1 + G_2$ has to
be kept fixed. The parameters are chosen in such a way that at $\theta
= 0$ reasonable values for the pion mass, chiral condensate and pion
decay constant are obtained~\cite{Frank:2003ve}. As we want to study
the effects of instantons, the parameter $c = G_2 / (G_1 + G_2)$ is
varied. In order for the model to have a stable ground state, this
parameter has to be between $0$ and $1/2$. Often $G_1$ and $G_2$ are
taken equal, i.e.\ $c=1/2$, which means that at $\theta=0$ the low
energy spectrum consists of $\sigma$ and $\bm{\pi}$ fields only. Using
the strange quark condensate it was argued in
Ref.~\cite{Frank:2003ve} that a realistic value of $c$ would be
around 0.2.
 
The ground state is obtained
by minimizing the effective potential, which is calculated in the
mean-field approximation. As usual, the ground state is described by
meson condensates, i.e.
\begin{align}
  \braket{\sigma} & = \braket{\bar \psi \lambda_0 \psi}, & \braket{\bm{a}_0} &  = \braket{\bar \psi \bm{\lambda} \psi}, \nonumber \\
  \braket{\eta} & = \braket{\bar \psi \lambda_0 i \gamma_5 \psi}, & \braket{\bm{\pi}}&  = \braket{\bar \psi \bm{\lambda} i \gamma_5 \psi}.
\end{align}

\section{Results}
All results presented here are for $\theta = \pi$. In the phase diagrams a
solid line denotes a first-order phase transition, a dashed line a
second-order, and a dotted line a cross-over.

We start with a discussion of the quark-mass dependence of the model. In
Fig.~\ref{phasediagram_mu0_md0} the phase diagram as a function of the
up and down quark masses is shown for $c = 0.4$.  The region where
$\braket{\eta}$ and $\braket{\pi^0}$ are non-zero corresponds to a
phase which violates CP invariance. Nonzero $\braket{\pi^0}$ only
occurs for nondegenerate quark masses, as it arises purely in
combination with explicit $SU(2)_V$ breaking. The asymptotes
of the phase transition are proportional to $c$. Similar results have been
obtained using a two-flavor chiral Lagrangian in Ref.~\cite{Tytgat:1999yx}.
However, in that calculation the phase transition at high $m_{u,d}$ (above 
the asymptotes) is absent. This is in contrast to the three-flavor 
chiral Lagrangian case studied by Creutz~\cite{Creutz:2003xu}, which does 
exhibit this second order phase transition. However, in that case the 
asymptotes correspond to the mass of the strange quark. 
 
To get more insight into the $c$ dependence of the SCPV phase, we show
in Fig.~\ref{phasediagram_m_c} the $(m, c)$ phase diagram for
degenerate quark masses ($m_u = m_d = m$). For every mass a critical
$c$ exists, above which SCPV occurs. From this, we can conclude that
the instanton interaction has to be strong enough w.r.t.\ the quark
masses in order to have SCPV. 

\begin{figure}[t]
  \begin{minipage}[t]{0.30\linewidth}
    \centering
    \includegraphics[width=\textwidth]{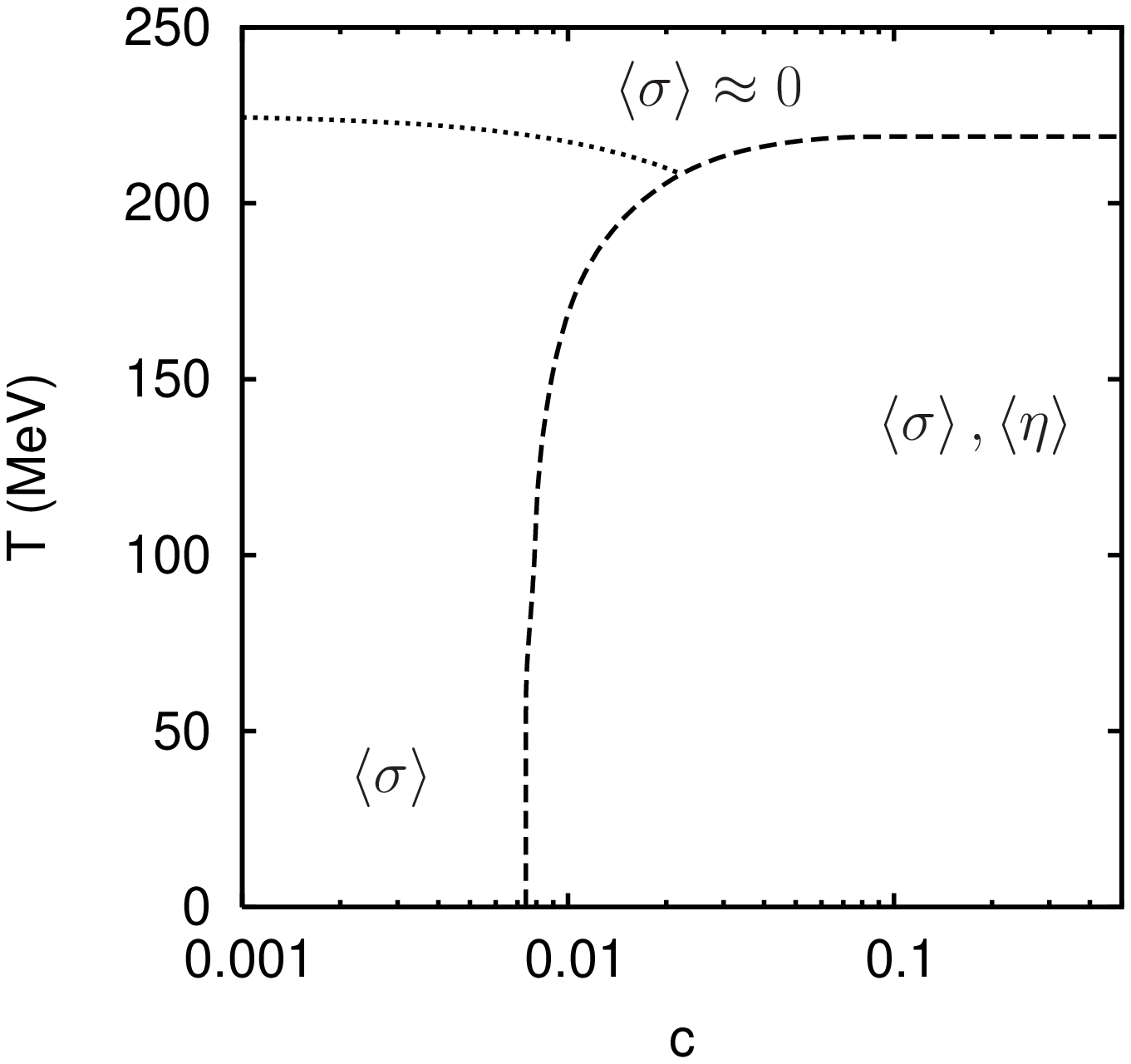}
    \caption{$(T, c)$ phase diagram.}
    \label{c_T_phasediagram}
  \end{minipage}%
  \hspace{0.05\linewidth}%
  \begin{minipage}[t]{0.3025\linewidth}
    \centering
    \includegraphics[width=\textwidth]{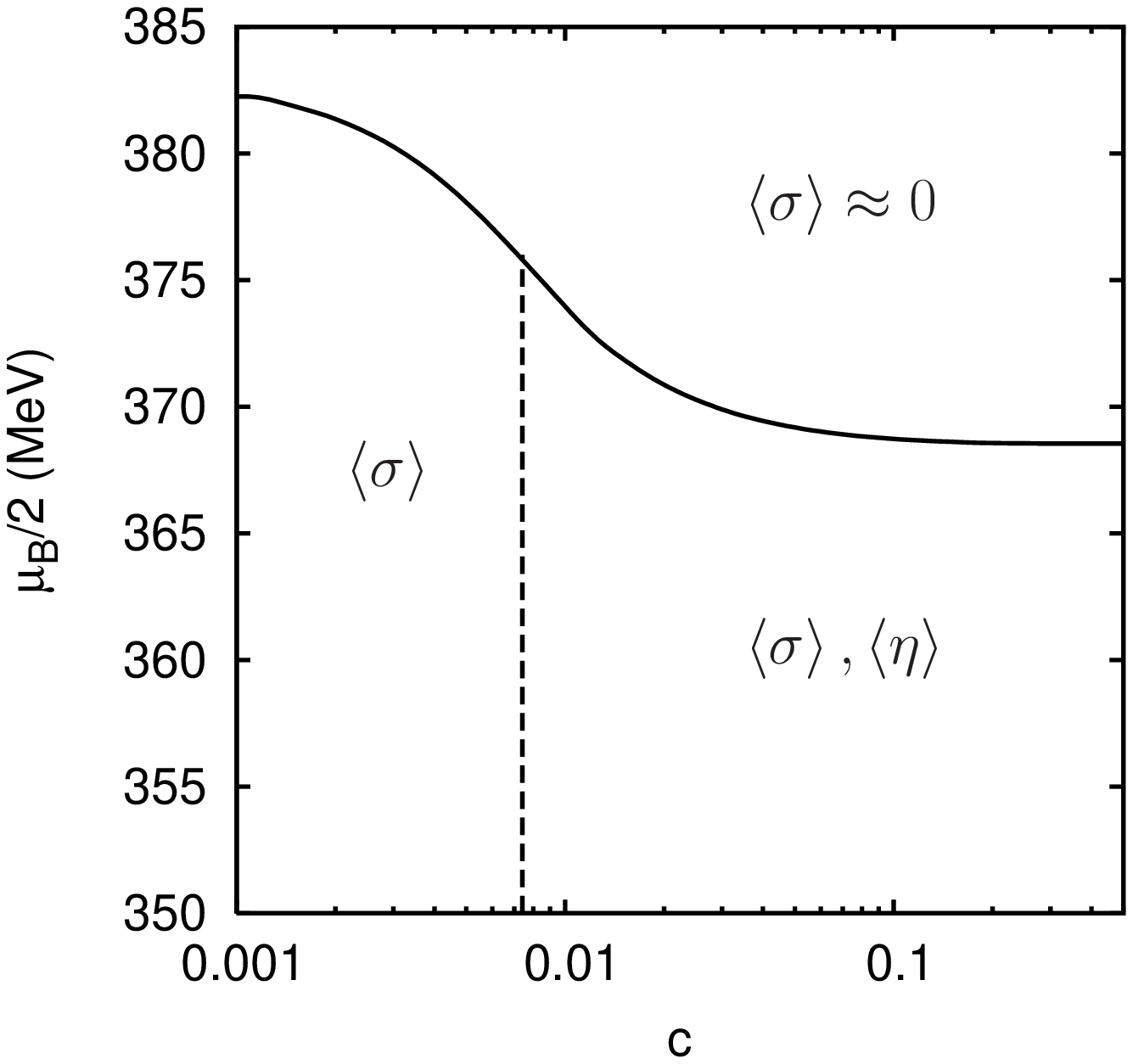}
    \caption{$(\mu_B, c)$ phase diagram.}
    \label{muB_c_phasediagram}
  \end{minipage}
  \hspace{0.05\linewidth}%
  \begin{minipage}[t]{0.30\linewidth}
    \centering
    \includegraphics[width=\textwidth]{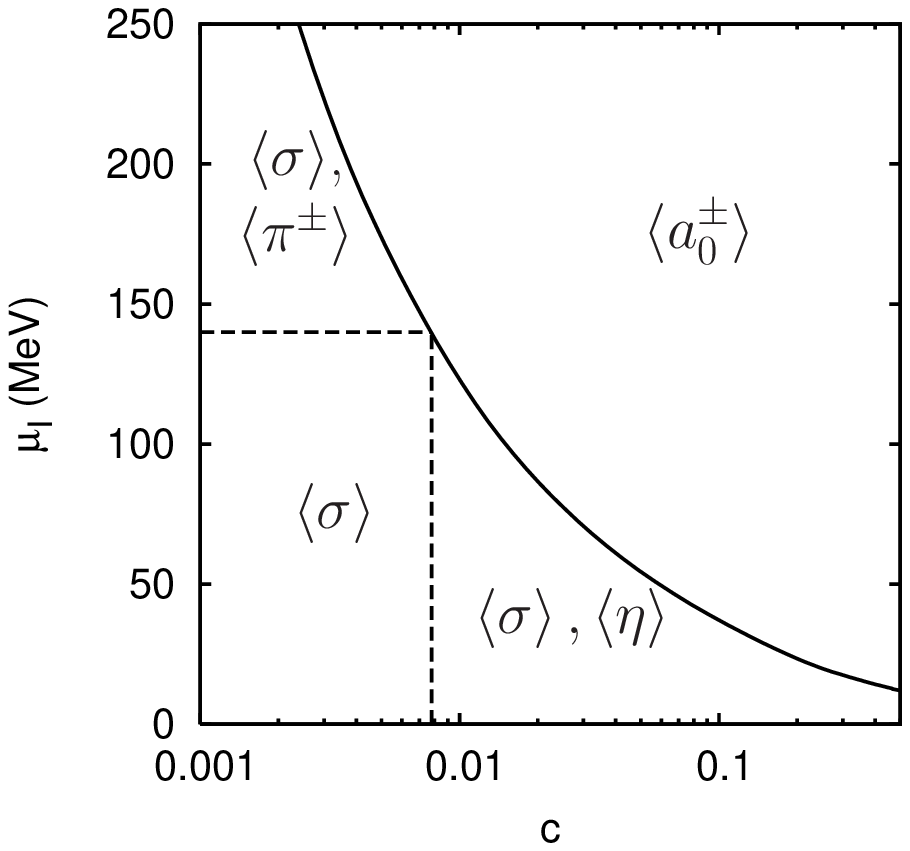}
    \caption{$(\mu_I, c)$ phase diagram.}
  \label{muI_c_phasediagram}
  \end{minipage}
\end{figure}

In Figs.~\ref{c_T_phasediagram} and \ref{muB_c_phasediagram} the
$(T,c)$ and $(\mu_B,c)$ phase diagrams are shown, where $T$ is the
temperature and $\mu_B = \mu_u + \mu_d$ is the baryon chemical
potential. When the temperature or the baryon chemical potential is
increased, the phase which violates CP invariance disappears, which
indicates that SCPV is a low energy phenomenon. It is therefore absent
in the deconfined phase at $\theta=\pi$. The CP-restoring phase
transition at high temperature is found to be of second order.  This
is in contrast to the results of Ref.\ \cite{Mizher:2008hf}, where a
first-order phase transition was found using a linear sigma model
coupled to quarks.

\begin{figure}[b]
   \begin{minipage}[t]{0.42\linewidth}
     \centering
     \includegraphics[width=\textwidth]{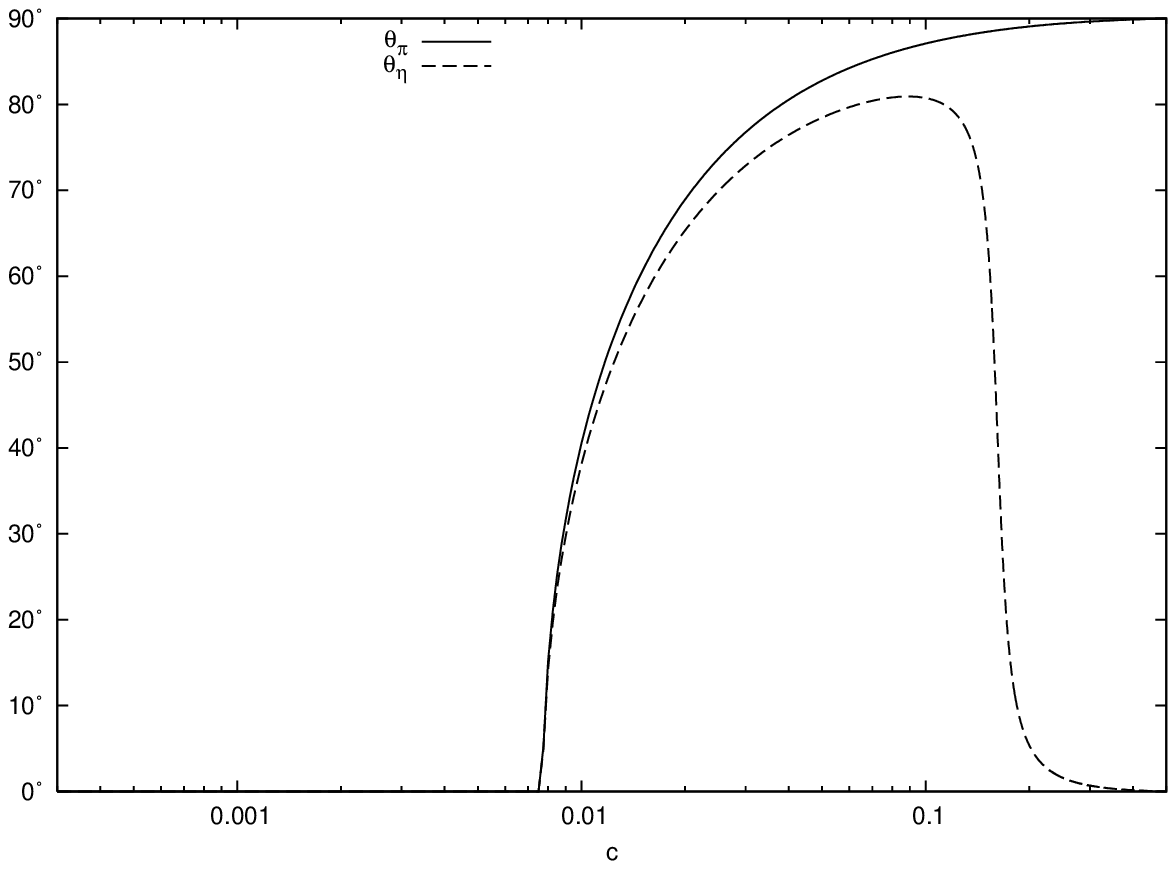}
     \caption{The $c$-dependence of the meson mixing angles at $\mu_I = 0$.}
     \label{c_dependence_mixing}
   \end{minipage}
  \hspace{0.16\linewidth}%
   \begin{minipage}[t]{0.42\linewidth}
     \centering
     \includegraphics[width=\textwidth]{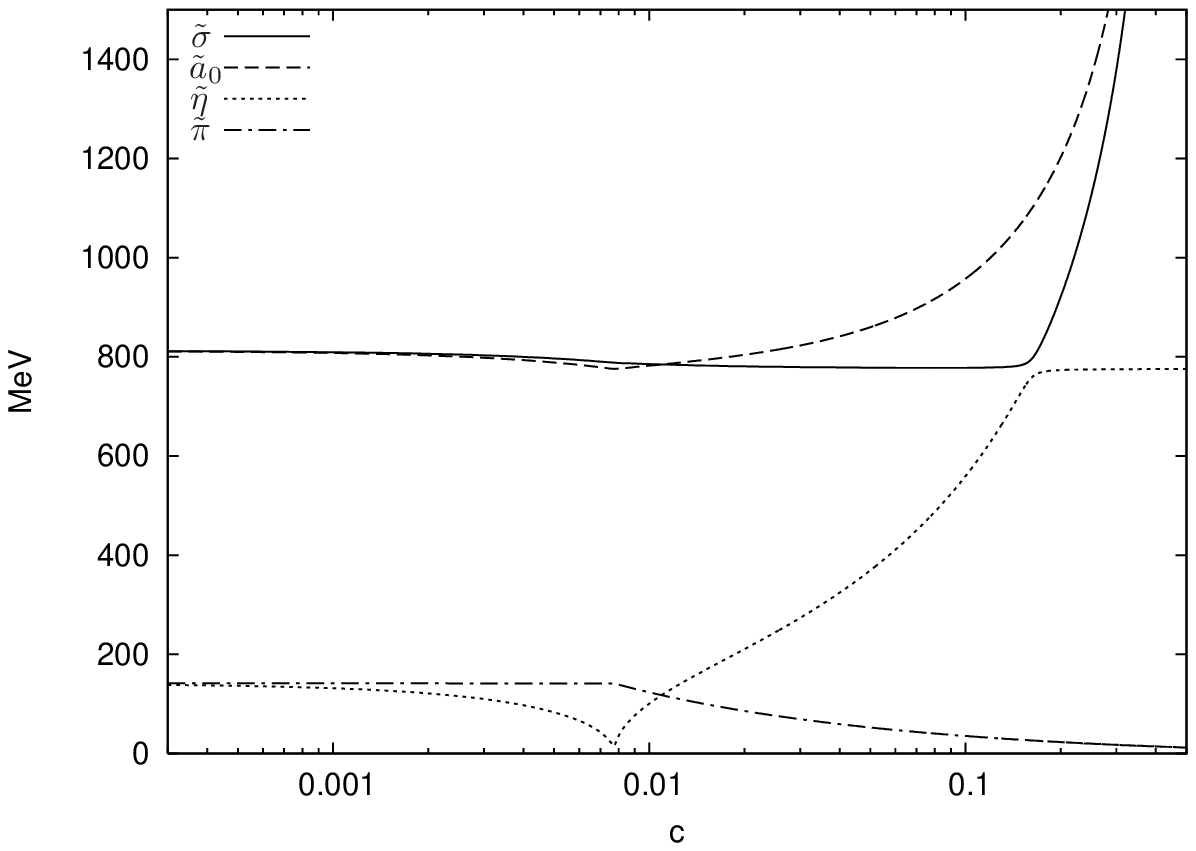}
     \caption{The $c$-dependence of the meson masses at $\mu_I = 0$.}
     \label{c_dependence_masses}
   \end{minipage}
\end{figure}
  
The final phase diagram which we discuss is the one as a function of
$c$ and the isospin chemical potential $\mu_I = \mu_u - \mu_d$, shown
in Fig.~\ref{muI_c_phasediagram}. Son and Stephanov~\cite{Son:2000xc} 
have shown that a (second-order)
phase transition to a charged pion condensed phase occurs 
at $\theta = 0$ when $\mu_I$ becomes larger than the vacuum pion mass. 
This condition still
holds at $\theta = \pi$ as long as there is no SCPV. However, for even larger 
$\mu_I$ there is second phase transition to a novel phase of 
$\bm{a}_0^\pm$-condensation, which is characteristic for $\theta =
\pi$. Moreover, this second phase transition is of first order.
Above the critical $c$, the charged pion condensate
does not occur for any value of $\mu_I$, only the charged $\bm{a}_0$ 
condensate. In order to discuss the condition for $\bm{a}_0^\pm$-condensation,
we have to calculate the mass eigenstates in the presence of SCPV.  
SCPV causes
mixing between parity partners, i.e.\ mass eigenstates are no longer CP
eigenstates. The mass eigenstates, denoted with a tilde, are defined as
\begin{align}
  \ket{\tilde \sigma}    & = \cos \theta_\eta \ket{\sigma} + \sin \theta_\eta \ket{\eta},
  & \ket{\tilde \eta}    & = \cos \theta_\eta \ket{\eta}   - \sin \theta_\eta \ket{\sigma}, \nonumber \\
  \ket{\tilde{\bm{a}}_0} & = \cos \theta_{\bm{\pi}} \ket{\bm{a}_0} + \sin \theta_{\bm{\pi}} \ket{\bm{\pi}},
  & \ket{\tilde{\bm{\pi}}} & = \cos \theta_{\bm{\pi}} \ket{\bm{\pi}} - \sin \theta_{\bm{\pi}} \ket{\bm{a}_0},
\end{align}
where $\theta_\eta$ and $\theta_{\bm{\pi}}$ are the mixing angles. The
states on the right-hand side are the usual states of definite parity.

The masses and mixing are calculated in the random phase approximation
\cite{Klevansky:1992qe}.  In Fig.~\ref{c_dependence_mixing} the $c$
dependence of the mixing angles is shown. When the CP violating
condensate turns on, the mixing angles become non-zero. In
Fig.~\ref{c_dependence_masses} the corresponding vacuum masses are
plotted. Comparing these masses with the phase diagram we can conclude
that the phase transition between the $\eta$ and the
$a_0^\pm$-condensate corresponds to the vacuum mass of $\tilde \pi$.
Fig.~\ref{c_dependence_mixing} shows that for the special case of
$c=1/2$ ($G_1=G_2$) the state $\ket{\tilde{\bm{\pi}}}$ is entirely
$\ket{\bm{a}_0}$. Furthermore, one can observe that when $c$
approaches $1/2$ the mass of the $\sigma$ and $\bm{a}_0$ fields go to
infinity which indicates that these fields decouple, as expected for $c = 1/2$.

In conclusion, the results presented here show that the phase structure of the
strong interactions at $\theta =\pi$ is more diverse than at $\theta=0$, 
thanks to spontaneous CP violation and the effects of instantons. 
We expect the presented two-flavor NJL-model results to remain valid
in the case of three flavors and when going beyond the mean-field
approximation, but this remains to be studied. Also, it would be
interesting if the results could in the future be compared to lattice QCD
results on the low-energy physics at $\theta = \pi$.

\section*{Acknowledgments}
We thank Harmen Warringa for his help regarding the
effective potential calculation.

\end{document}